\documentclass[lettersize,journal]{IEEEtran}
\def\BibTeX{{\rm B\kern-.05em{\sc i\kern-.025em b}\kern-.08em
    T\kern-.1667em\lower.7ex\hbox{E}\kern-.125emX}}

\usepackage{algorithm}
\usepackage{algorithmic}
\usepackage{cite}
\usepackage{amsmath,amssymb,amsfonts}
\usepackage{algorithmic}
\usepackage{graphicx}
\usepackage{textcomp}
\usepackage{xcolor}
\usepackage{breqn,xspace}
\usepackage{bm}
\usepackage{multirow}
\usepackage{threeparttable}
\usepackage{mathrsfs}
\usepackage{pifont}
\usepackage{caption}
\usepackage{subfig}
\usepackage{lipsum,multicol}
\usepackage{makecell}
\usepackage{booktabs}

\newcommand{\name}{Sums\xspace}

\ifodd 0
\definecolor{darkgreen}{RGB}{0,200,0}
\newcommand{\rev}[1]{{\color{blue}#1}} 
\newcommand{\newrev}[1]{{\color{red}#1}} 
\newcommand{\needrev}[1]{{\color{darkgreen}#1}} 
\else
\newcommand{\rev}[1]{#1}
\newcommand{\newrev}[1]{#1} 
\newcommand{\needrev}[1]{#1} 
\fi

\begin{document}
\title{\name: Sniffing Unknown Multiband Signals under Low Sampling Rates}

\author{Jinbo~Peng, Zhe~Chen,~\IEEEmembership{Member,~IEEE}, Zheng~Lin, Haoxuan~Yuan, Zihan~Fang, Lingzhong~Bao, Zihang~Song,~\IEEEmembership{Member,~IEEE}, Ying~Li~\IEEEmembership{Member,~IEEE}, Jing~Ren~\IEEEmembership{Member,~IEEE}, and~Yue~Gao,~\IEEEmembership{Fellow,~IEEE} 
\IEEEcompsocitemizethanks{
\IEEEcompsocthanksitem
J. Peng, Z. Chen, H. Yuan, L. Bao and Y. Gao are with the School of Computer Science, Fudan University, Fudan University, Shanghai 200438, China (email: jbpeng22@m.fudan.edu.cn; zhechen@fudan.edu.cn; hxyuan22@m.fudan.edu.cn; lzbao22@m.fudan.edu.cn; gao.yue@fudan.edu.cn).
\IEEEcompsocthanksitem
Z. Lin is with the Department of Electrical and Electronic Engineering, University of Hong Kong, Pok Fu Lam, Hong Kong, China (e-mail: linzheng@eee.hku.hk).
\IEEEcompsocthanksitem
Z. Fang is with the Department of Computer Science, City
University of Hong Kong, Kowloon, Hong Kong, China (e-mail: zihanfang3@cityu.edu.hk).
\IEEEcompsocthanksitem
Z. Song is with the Department of Engineering, King's College London, Strand, London, WC2R 2LS, United Kingdom (email: zihang.song@kcl.ac.uk). 
\IEEEcompsocthanksitem
Y. Li and J. Ren are with Peng Cheng Laboratory, Shenzhen, China (email: liy02@pcl.ac.cn; renj@pcl.ac.cn). 
\IEEEcompsocthanksitem
J. Ren is also with University of Electronic Science and Technology of China, Chengdu, China (email:renjing@uestc.edu.cn).
}
\thanks{}}

%
%

\markboth{Journal of \LaTeX\ Class Files,~Vol.~14, No.~8, August~2015}%
{Shell \MakeLowercase{\textit{et al.}}: Bare Advanced Demo of IEEEtran.cls for IEEE Computer Society Journals}

\maketitle

\begin{abstract}
Due to sophisticated deployments of all kinds of wireless networks~(e.g., 5G, Wi-Fi, Bluetooth, LEO satellite, etc.), multiband signals distribute in a large bandwidth~(e.g., from 70\!~MHz to 8\!~GHz). Consequently, for network monitoring and spectrum sharing applications, a sniffer for extracting physical layer information, such as structure of packet, with low sampling rate~(especially, sub-Nyquist sampling) can significantly improve their cost- and energy-efficiency.
However, to achieve a multiband signals sniffer is really a challenge. 
To this end,
we propose \name, a system that can sniff and analyze multiband signals in a blind manner.
Our \name takes advantage of hardware and algorithm co-design, multi-coset sub-Nyquist sampling hardware, and a multi-task deep learning framework. The hardware component breaks the Nyquist rule to sample GHz bandwidth, but only pays for a 50\!~MSPS sampling rate. \newrev{Our multi-task learning framework directly tackles the sampling data to perform spectrum sensing, \newrev{physical layer protocol recognition}, and demodulation for deep inspection from multiband signals.}
Extensive experiments demonstrate that \name achieves higher accuracy than \newrev{the state-of-the-art baselines} in spectrum sensing, modulation classification, and demodulation. As a result, our \name can help researchers and end-users to diagnose or troubleshoot their problems of wireless infrastructures deployments in practice.
\end{abstract}


\section{Introduction}\label{sec:intro}

\IEEEPARstart{W}{ireless} networks have been deployed everywhere in our daily life. From outdoor to indoor, there are various kinds of wireless technologies such as satellite communication, LTE, Wi-Fi, Bluetooth, etc. that occupied different bands. 
Due to such complicated deployment of wireless networks, any oversight in resource management easily results in overlapping spectra, leading to mutual electromagnetic interference and significant deterioration in communication performance.
For example, different LEO satellites launched by various parties may transmit signals to ground stations simultaneously at the same location, and thus interfere with each other~\cite{spacexKuiper2021}. The throughput and latency of Wi-Fi networks may be severely affected by interference from other industrial medical (ISM) band devices~\cite{gummadi2007understanding}.
\newrev{This situation inevitably calls for a powerful sniffer for wireless network telemetry and spectrum monitoring~(e.g., 1\!~GHz bandwidth). In particular, multiband wireless signal monitoring is significant for spectrum-sharing applications (e.g., coexistence of non-terrestrial and terrestrial networks~\cite{alexandre2020coexistence}).}


\begin{figure}[t!]
  \centering
  \includegraphics[width=0.49\textwidth]{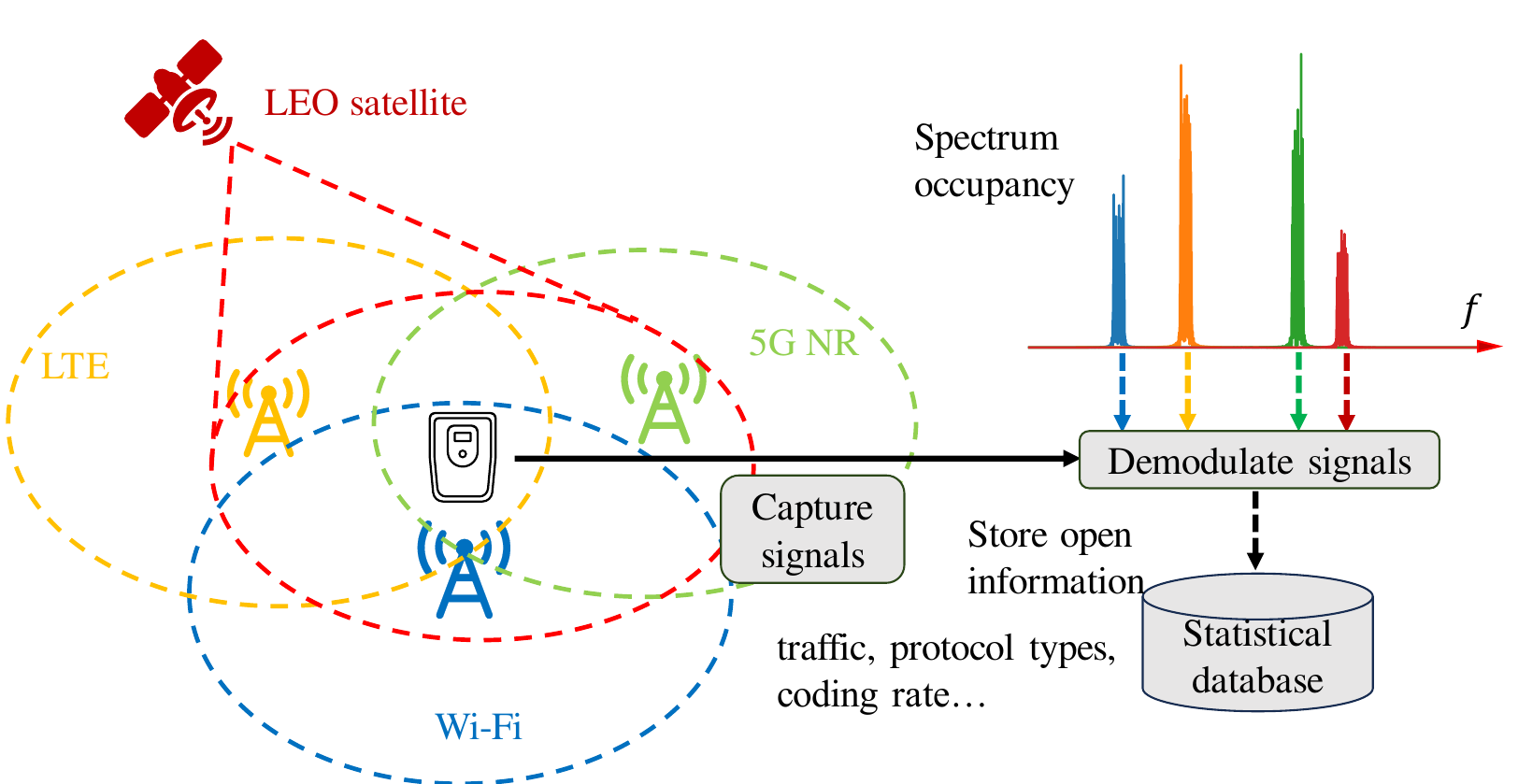}
      \caption{An example scenario for unknown multiband signals monitoring in a wideband spectrum.}
  \label{fig:teaser}
    \vspace{-5ex}
\end{figure}

\newrev{
A wireless sniffer not only captures the features of the spectrum, but also extracts more detailed wireless signal features~(i.e., modulation and coding scheme, type of frame, channel quality, etc.) from wireless signals. The information of these features extracted from the headers of physical layer frames can be applied in applications such as deep packet inspection~\cite{deri2014ndpi}, network diagnostics~\cite{liu2014self}, and spectrum sharing.}
\rev{As illustrated in Fig.~\ref{fig:teaser}, the sniffer blindly captures four different types of signals in a wideband spectrum, and then digs out \newrev{physical layer} information}. \newrev{By analyzing open headers of physical layer protocols, we can identify errors, anomalies or misconfigurations in the communication infrastructure, thus facilitating operators to pinpoint the source of problems and take appropriate actions. Moreover,}
The sniffer enables the accommodation and coexistence of new and emerging wireless technologies with existing systems, while also allowing for the detection of anomalous signals.

\newrev{
Recalling that most of existing sniffers~(e.g., Wireshark~\cite{wireshark2024}, Kismet~\cite{kismet2024}, and tcpdump~\cite{tcpdump2024}), they always need a full analog-to-digital converter (ADC) sampling rate. Due to the larger and larger bandwidth used in future wireless networks, the cost of a radio receiver with high-speed sampling rate ADC is really expensive for sniffers. Therefore, recent state-of-the-art work Swirls~\cite{gao2023swirls} proposes sub-Nyquist sniffing solutions to reduce the cost of radio receiver. However, it only focus on a specific physical layer protocol, such as Wi-Fi, and its certain single band signals. It cannot tackle multiband signals~(shown in Fig.~\ref{fig:teaser}), and multiple physical layer protocols, simultaneously under a low sampling rate~(a.k.a sub-Nyquist).} The naive way to decode multiband signals is to scan sequentially the whole bands via channel hopping generated by sub-Nyquist sampling rate radio~\cite{guddeti2019sweepsense, shi2015beyond}. Nevertheless,
our observations indicate that the carrier frequency switching period has an impact on both spectrum sensing and blind demodulation, as discussed in Sec.~\ref{sec:motivation_background}. Faster switching periods yield better spectrum sensing performance but degrade blind demodulation performance, and vice versa. It appears that such a dilemma cannot be totally solved by current work.

In this paper, we propose, design, and implement \name, \newrev{a monitoring sniffer that passively and \needrev{blindly decodes unknown multiband signals in a large wideband spectrum.} To the best of our knowledge, \name is the first sniffing system which \needrev{decodes multiband signals} under a low sampling rate.}  
However, to achieve that system, we still face three core challenges in the following:
\begin{itemize}
    \item Directly sampling multiband signals requires expensive wideband radio~(at least equal to Nyquist rate using IQ sampling).  
    \item In order to sniff that \newrev{multiband signals}, due to frequency switching period, narrowband radio based solutions meet a dilemma situation for accurate spectrum sensing and blind demodulation.
    \item \newrev{Multiband signals may have different physical layer protocols. They are mixed together under a sub-Nyquist sampling rate, and hard to distinguish blindly for decoding.}
\end{itemize}
To solve the above challenges, we propose, design and implement a hardware and algorithm co-designed blind monitoring system, named \name which can sniff multiband signals in a wideband spectrum. Instead of conventional scanning approaches via a narrowband radio, we employ 8 low speed ADCs~(operating at 50~\!MSPS) based on multi-coset sampling method~\cite{mishali2009blind, venkataramani2001optimal} to capture sub-Nyquist samples across a wideband spectrum. We implement a data pipeline in FPGA to quickly transfer sub-Nyquist samples to a host PC via PCIE bus. To efficiently dig out information from the sub-Nyquist samples, inspired by recent foundation models, such as ChatGPT, we design a flexible and compact multi-task learning framework based on Transformer encoder layers. Our multi-task learning framework models the whole blind monitoring process in four tasks: spectrum sensing, modulation classification, \newrev{physical layer protocol recognition} and blind demodulation. The neural network outputs of spectrum sensing are used to guide training of \newrev{the other three tasks.}
\needrev{\name enables the simultaneous processing of multiple heterogeneous signals using a light-weight and efficient network architecture.} The hardware and algorithm co-design of \name not only improves monitoring coverage of wideband spectrum, but also enables rapid demodulation of unknown multiband signals.

We briefly summarize our contributions in the following.
\rev{
\begin{itemize}
    \item We design a hardware and algorithm co-designed system to sniffer unknown multiband signals in a large wideband spectrum.
    \item We develop a multi-coset sampling hardware using only 8 narrowband radios with 50~\!MHz bandwidth to sample wideband spectrum with GHz bandwidth.
    \item According to above sub-Nyquist samples, we design an end-to-end multi-task learning framework for unknown multiband signals analysis.
    \item \needrev{We propose a light-weight and efficient network architecture to handle different types of signals}
    \item We implement a prototype of \name, and give extensive evaluations to demonstrate \name outperforms the state-of-the-art solutions~\cite{tropp2005simultaneous,huynh2020mcnet}.
\end{itemize}
}

In the following, Sec.~\ref{sec:motivation_background} motivates the design of \name by revealing the problem of current blind monitoring solutions. Related works are discussed in Sec.~\ref{sec:related_work}. Sec.~\ref{sec:system_design} describes the system design of \name. Sec.~\ref{sec:implementation} and Sec.~\ref{sec:evaluation} present system implementation, and evaluation results, respectively. In the end, we conclude this paper in Sec.~\ref{sec:conclusion}.

\section{Motivation and Background} \label{sec:motivation_background}



\subsection{Multiband Signals Sampling} \label{ssec:mss}
Multiband signals mean wideband signals composed of narrowband signals with different carrier frequencies, and their corresponding bandwidths in the air shown in Fig.~\ref{fig:teaser}. Those narrowband signals always distribute in a large wideband spectrum. Generally, there are two methods to capture all of them. One is to leverage a wideband radio~(e.g., 1\!~GHz)~\cite{goodson2017multi}, and the other one is to sweep all possible carrier frequencies via a narrowband radio~\cite{guddeti2019sweepsense, shi2015beyond}. For the first method, a high speed ADC is leveraged to capture all information from the spectrum. Although conventional blind estimation algorithms~\cite{hanna2021signal, o2017introduction, swami2000hierarchical} only consider single-band signal, each single-band signal of the multiband signal can be distinguished easily, and processed independently. However, in practice, since the spectrum range of multiband signals is large, current commercial-off-the-shelf wideband radio does not achieve enough bandwidth for them\footnote{Even there exists ADC with more than 1\!~GSPS, it is extremely expensive.}. Consequently, the frequency sweeping method is designed to reduce the cost of wideband radios, and it is successful to apply in spectrum sensing task via well-designed sweeping algorithms~\cite{guddeti2019sweepsense, shi2015beyond}. But spectrum sensing is the first step for a blind monitor, and it further struggles to dive deep into the signals. In the following section, we explore the limitation of the sweeping method in blind estimation tasks.  

\begin{figure}[t]
  \centering
  \subfloat[Switching periods impact on modulation classification]{
    \includegraphics[width=0.45\linewidth]{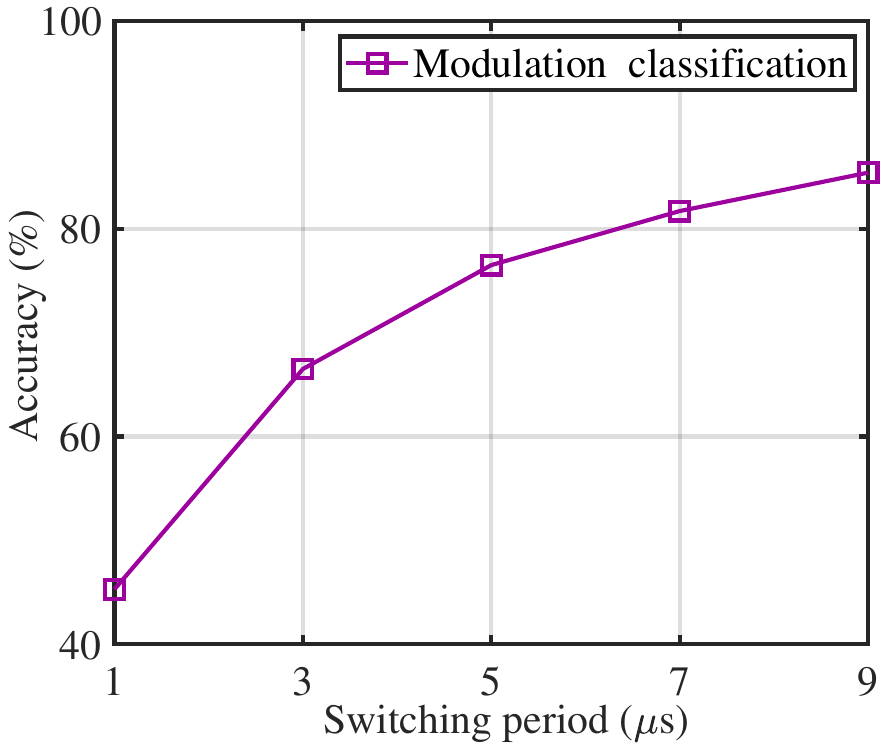}
    \label{sfig:period_amc}
  } \hspace{0.01\linewidth}
  \subfloat[Switching periods impact on spectrum sensing]{
    \includegraphics[width=0.45\linewidth]{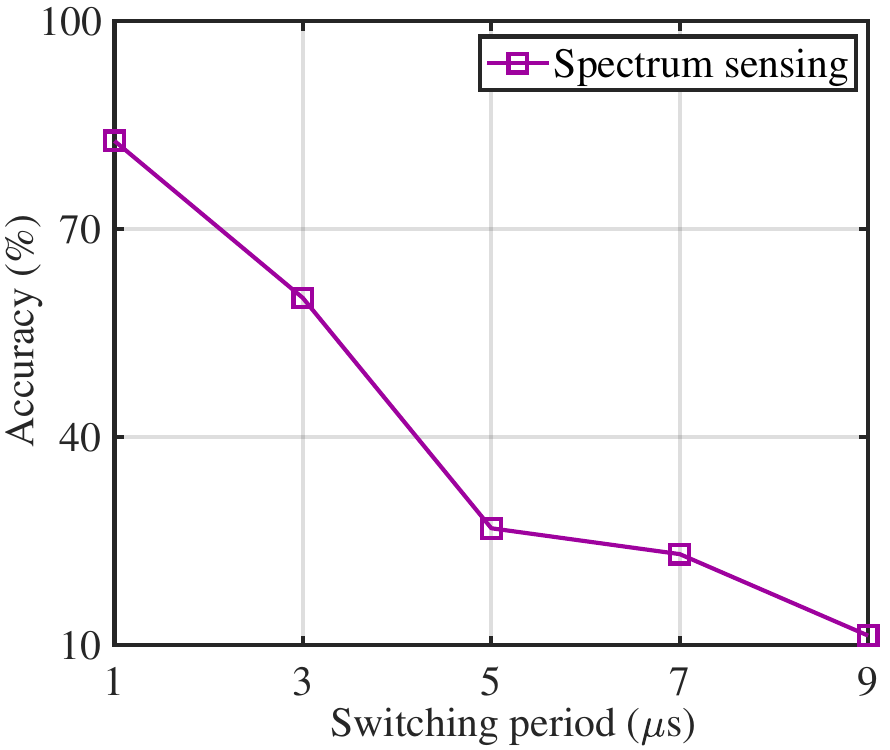}
    \label{sfig:period_ss}
  }\\  
  \subfloat[Carrier frequency offsets impact on demodulation]{
    \includegraphics[width=0.45\linewidth]{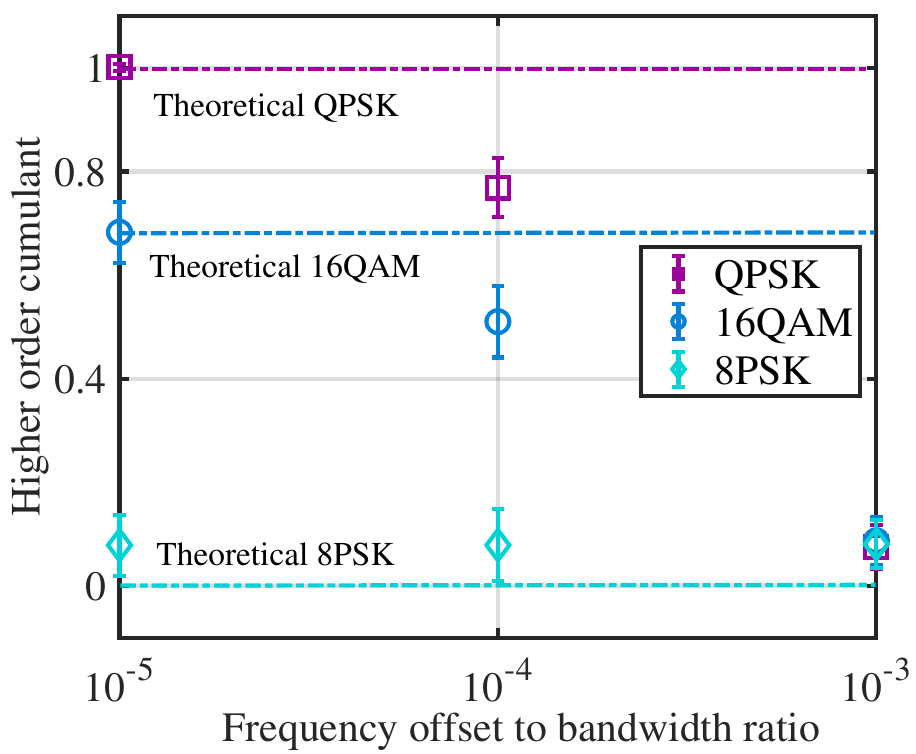}
    \label{sfig:cumulant_freq_offset}
  } \hspace{0.01\linewidth}
  \subfloat[Different symbol lengths impact on demodulation]{
    \includegraphics[width=0.45\linewidth]{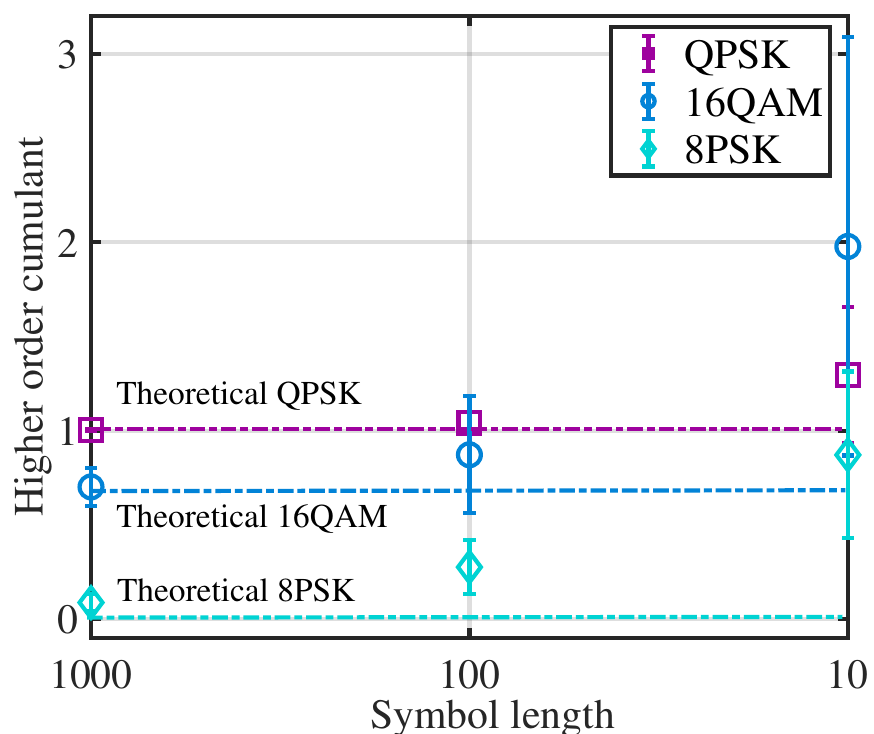}
    \label{sfig:cumulant_symb_len}
  }
  \caption{Mutual effects between spectrum sensing and blind demodulation.}
  \label{fig:motivation}
\end{figure}

\subsection{Frequency Sweeping Limitation}
For a blind monitor, the final goal is to demodulate signals. 
Before blind demodulation, blind modulation classification plays an important role, due to its wrong results making demodulation certain to fail. Consequently, in the following, we study blind modulation classification instead of blind demodulation. 
In our measurement experiments, for convenience, we employ a classic spectrum sensing algorithm~\cite{tropp2005simultaneous} and a blind modulation classification algorithm~\cite{swami2000hierarchical} to study how the frequency sweeping method affects their performance. This simplification does not compromise the generality, as both deep learning-based and model-based solutions face the same challenge in this context. For the sweeping method, the switching period means time of duration between two hopping frequency points.
We set the total bandwidth to 1~\!GHz, and randomly generate 4 multiband signals with different time durations distributed in a time-frequency spectrum. Then, we increase the switching period from 1 to 9~\!$\mu$s, and plot the accuracy of blind modulation classification and spectrum sensing in Fig.~\ref{sfig:period_amc} and Fig.~\ref{sfig:period_ss}, respectively. Apparently, their performance is mutually affected due to switching periods. 
Reducing the switching period in spectrum sensing enables a faster sweep across the entire spectrum, thereby enhancing the ability to accurately detect multiband signals with shorter duration.
On the other hand, blind modulation classification needs a longer switching period, since the algorithm designed by~\cite{swami2000hierarchical} relies on a sufficient number of samples to extract relevant features.
We also display the higher-order cumulants to represent the capability of modulation classification. In Figure~\ref{sfig:cumulant_freq_offset}, larger carrier frequency offsets due to spectrum sensing errors, and in Figure~\ref{sfig:cumulant_symb_len}, shorter signal duration, result in classification failures as indicated by the deviation of the cumulant values from the theoretical values. Therefore, it is non-trivial to design a blind monitor, that jointly solve spectrum sensing, modulation classification and demodulation problems.

\rev{
\subsection{Heterogeneous Signals}
In addition to the challenges in signal estimation, designing a concise and flexible system capable of simultaneously handling multiple types of signals is also highly challenging. Traditional signal processing pipelines involve complex modules such as filtering, signal estimation, offset compensation, and more. Although deep learning-based demodulation systems have simplified the process of signal parameter estimation, they still face difficulties in accommodating diverse signal types. For instance, input signals can vary in duration, making it challenging for conventional Convolutional Neural Network (CNN) to handle inputs of different lengths. Moreover, signals with different bandwidths have varying numbers of transmitted symbols, necessitating the dynamic adjustment of output lengths by neural networks. Furthermore, when confronted with inputs of multiband signals, the model needs to differentiate mixed narrowband signal for subsequent signal analysis. Consequently, the model is required to analyze multiple heterogeneous signals from a mixed input sequence, posing challenges to the structural design of the signal analysis network.
}

\section{System Design} \label{sec:system_design}

\begin{figure*}[t!]
  \centering
  \subfloat[Multi-coset sub-sampling and the embedding of samples. Samples from ADCs are embedded into row vectors]{
      \includegraphics[width=0.75\textwidth]{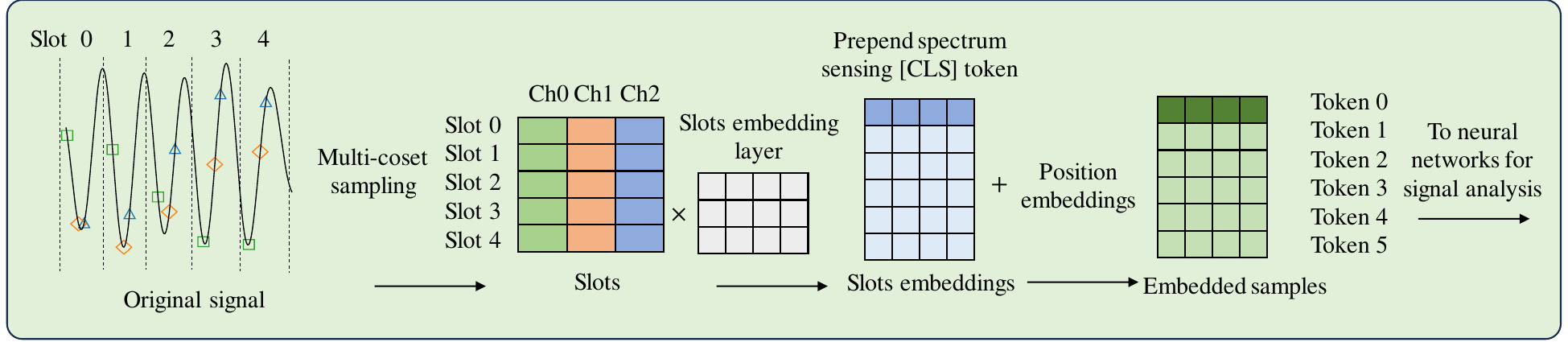}
      \label{fig:embedding}
  } \\
  \subfloat[The multi-task learning framework based on Transformer encoders for signal analysis. For conciseness, each small square in this figure represents a row vector of dimension $d_{model}$.]{
      \includegraphics[width=0.75\textwidth]{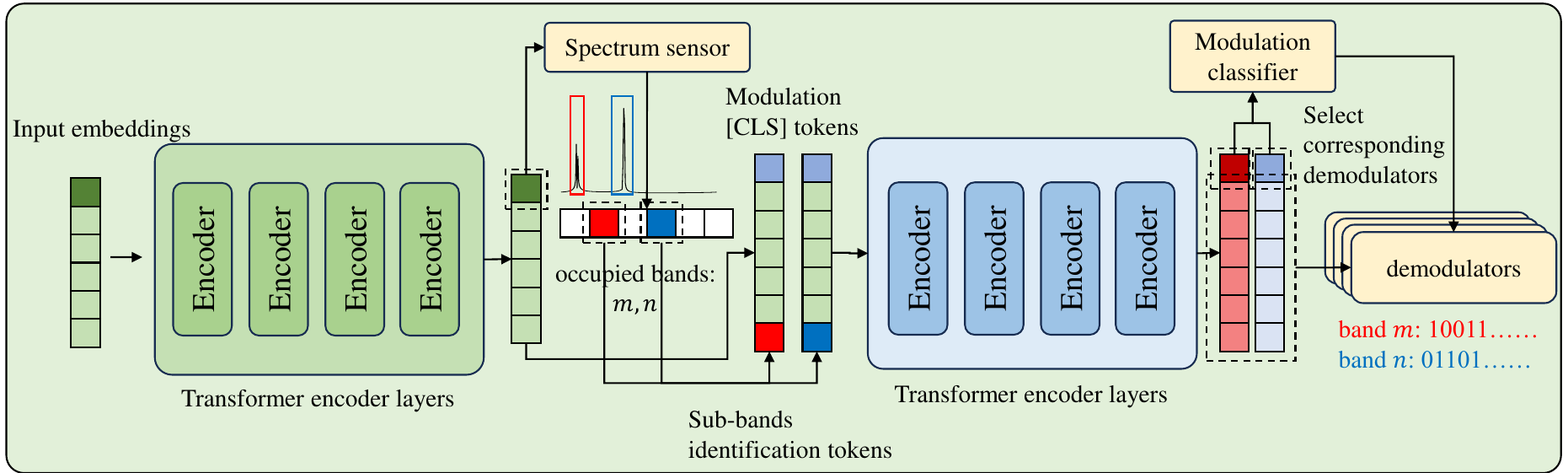}
      \label{fig:network}
  }
  \caption{\rev{The architecture of \name comprises two main components: the multi-coset sub-sampling to capture wide-spectrum signal efficiently and the signal analysis network for spectrum sensing, modulation classification and blind demodulation.}}
  \label{fig:architecture}
\end{figure*}

\rev{
\name is comprised of the multi-coset sub-sampling circuit (Sec.~\ref{ssec:multi-coset}) and the signal analysis network (Sec.~\ref{ssec:san}). The architecture of \name is illustrated in Fig.~\ref{fig:architecture}. The multi-coset sub-sampling circuit is designed for efficient multiband signals sampling in a large wideband spectrum, and the network is a Transformer multi-task learning framework that directly processes samples from multi-coset sub-sampling circuit. This network performs signal analysis tasks needed for multiband signals sniffing, including spectrum sensing, multiband modulation classification, physical layer protocol recognition and demodulation\footnote{\name focuses on demodulation, since current protocol analyzers, such as Wireshark, tcpdump, etc., are capable of decoding physical layer bit streams after demodulation.}.
}

\subsection{Multi-coset Sub-sampling Hardware} \label{ssec:multi-coset}

\begin{figure}[t!]
  \centering
  \includegraphics[width=0.42\textwidth]{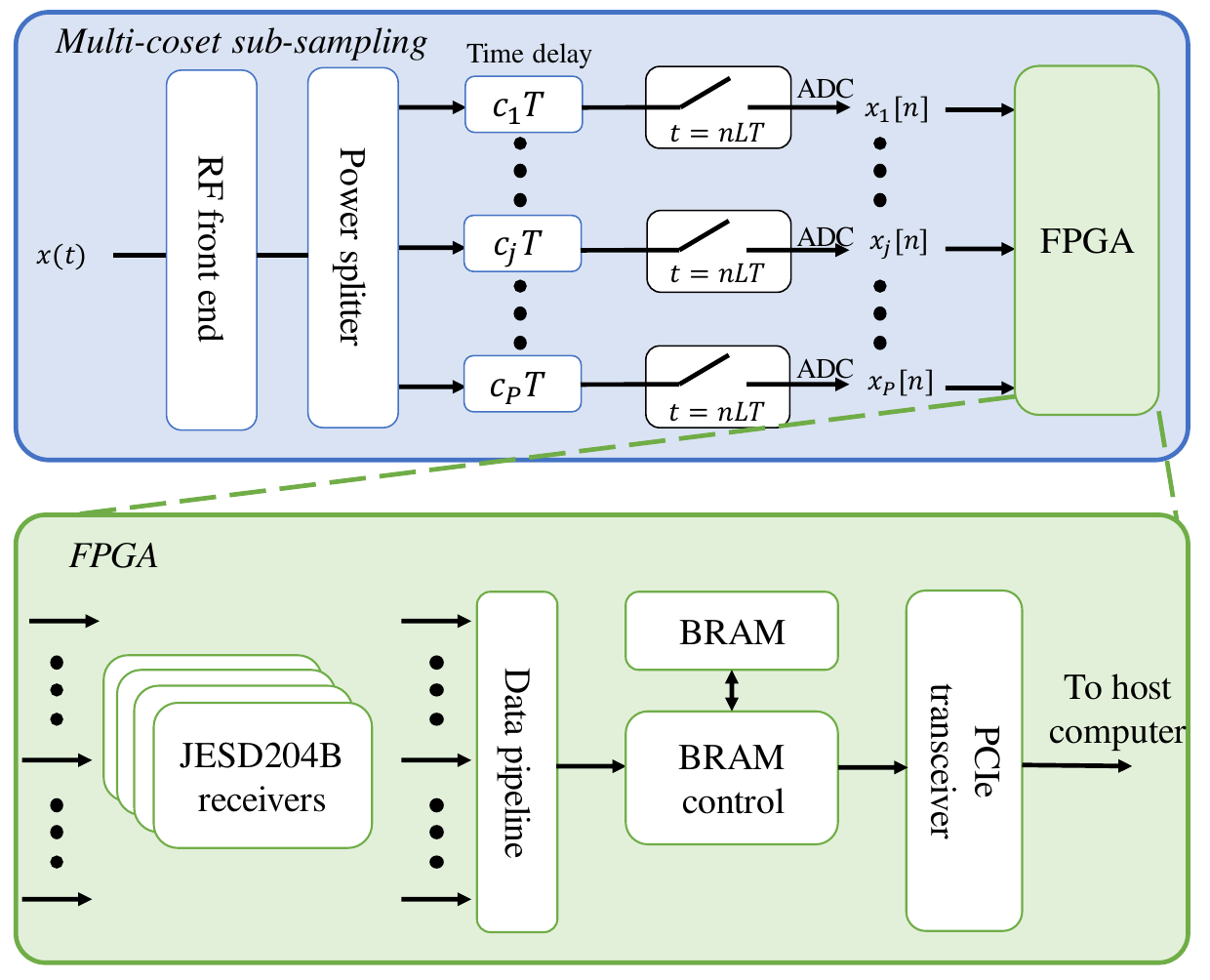}
  \caption{Multi-coset sub-sampling hardware configuration.}
  \label{fig:multicoset}
\end{figure}

This section presents the hardware design of multi-coset sub-sampling in the \name framework. For our hardware configuration, the original signal is first assigned to multiple ADCs for parallel sampling, and then sampled data is received and processed using a well-designed FPGA module, which lays the foundation for spectrum sensing and blind demodulation in subsequent data analysis networks. 

As illustrated in Fig.~\ref{fig:multicoset}, the original signal first reaches the RF front end, and then goes through an equal-power splitter so that each ADC is distributed a signal with the identical power level as the original signal. Unlike existing methods that use a single ADC to directly perform frequency-hopping scanning and sampling across a wideband spectrum, the multi-coset sampling approach utilizes multiple ADCs with unique time delays to perform parallel sampling of the signal. This allows sampling signals with  wider spectrum than a single ADC can handle, enabling higher precision in signal reconstruction and, consequently, more accurate analysis and interpretation of complex signal data. $x(t)$ is band-limited original signal within the range $\left[ {0, B} \right]$, where $B$ represents the maximum frequency of the band-limited signal. The Nyquist sampling interval is denoted by $T=\frac{1}{2B}$. In this multi-coset sub-sampling approach, the signals are sampled by $P$ ADCs in parallel and each ADC is configured with same sampling interval $LT$ ($P < L$),  where $L$ is a constant representing the ratio of the ADC sampling interval to the Nyquist sampling interval. Consequently, the sampled sequence of the $j$-th ADC is given by
\begin{equation}
{x_j}\left[ n \right] = x\left( {nLT + {c_j}T} \right),\quad n = 1,2,...,N,
\label{equ:cs}
\end{equation}
where $c_j$ denotes a non-negative integer satisfying $0 \le {c_j} < L$ and ${c_j} \ne {c_k}\left( {\forall k \ne j} \right)$ and $N$ is the number of sampled data for each ADC.

The recoverability of the original signal from sub-Nyquist samples indicates that sub-Nyquist samples indeed holds sufficient useful information about the original signal. Therefore, instead of pre-processing of the ADC-acquired sub-Nyquist samples (e.g., signal recovery), we directly send the information-rich sub-Nyquist samples to host PC via a well-designed FPGA module for spectrum sensing and blind demodulation. For the designed FPGA module, we first deploy JESD204B receivers to receive sub-Nyquist samples from multiple ADCs, and then perform parallel-to-serial conversion by data pipeline. After that, the converted data is temporarily cached in BRAM and finally transmitted  to host PC via PCIe bus. The host PC then will utilize a signal analysis network to extract features from these received sub-Nyquist samples and conduct further analysis on them, which is elaborated in the following section.

\subsection{Signal Analysis Network} \label{ssec:san}


In this section, we present the design details of the signal analysis network in \name,  as depicted in Fig.~\ref{fig:architecture}. Deep learning (DL), utilizing NNs, can extract features from massive wireless communication data that are superior to manually constructed ones~\cite{lin2023efficient,chen2021rf,yuan2023graph,lin2023fedsn,zheng2023autofed,lin2023split,qiu2024ifvit,lin2023pushing,fang2024automated,lin2024adaptsfl}. Motivated by this, we propose a deep multi-task learning framework based on the Transformer encoder layers for analyzing the sub-Nyquist samples acquired by multi-coset sub-sampling hardware.
\rev{
The design of the signal analysis network has the following considerations. Firstly, it employs Transformer, slots tokens and sample embeddings to handle variable-length inputs and outputs in a flexible manner. Secondly, recognizing that the demodulation process for different modulated signals shares substantial similarities, with the main distinction being the constellation mapping, we aim to maximize the utilization of a shared backbone network across various signal types. Lastly, identification tokens are introduced to guide the network's attention towards specific narrowband signals, enabling network structure reusability.
}

\rev{
Our signal analysis network consists of four primary modules. The first module constructs the samples into a sequence of continuous vector token embeddings. The remaining three modules are used for concrete signal analysis tasks. The samples embeddings are utilized as input and Transformer encoders are employed as the backbone of the network. They are responsible for extracting signal features and are powerful in finding long-range signal correlations. We use Fully Connected Layers (FCL) as the branches to map the extracted features from the backbone to produce the outputs of specific tasks, which are then fed back to the backbone to focus on downstream feature extraction. The second module is dedicated to spectrum sensing, while the third module focuses on signal classification for modulation classification, physical layer protocol recognition. The fourth module maps the high-dimensional features to symbols for demodulation.
}
In the following, we will provide corresponding design details for each module.

\subsubsection{Samples Embedding Module}
\label{sec:system_design:san:embedding}

\rev{
Traditional Transformer is capable of handling inputs of variable lengths. It accepts a sequence of vectors with a dimension of $d_{model}$ as its input. The first module of the signal analysis network is responsible for dividing the multi-coset samples into slots and constructing slots embeddings to serve as the input of the Transformer.

Referring to Eq.~\eqref{equ:cs}, the multi-coset sampling results by $P$ ADCs can be seen as an $N \times P$ complex matrix $X$, where $X_{n, j} = x_j[n] = x(nLT + c_j T)$. We consider each signal segment of duration $LT$ as a slot, so each ADC in the multi-coset sampling performs one sampling within a slot. Each row of $X$ represents the sampling results of all the low-speed ADCs within a slot, as shown in Fig.~\ref{fig:embedding}.

We unfold the complex matrix $X$ into a real matrix $X' \in \mathbb{R}^{N \times 2P}$, where $X'_{n, 2j+1}$ and $X'_{n, 2j+2}$ represent the real and imaginary parts, respectively, of the $j$-th ADC's sampling in the $n$-th slot. $X'$ is then passed through an FCL to obtain $N$ vectors of dimension $d_{model}$, which are slot embeddings. Similar to the [CLS] token in BERT~\cite{devlin2018bert} and ViT~\cite{dosovitskiy2020image}, we prepend a learnable spectrum sensing [CLS] token to the slot embeddings. This token integrates and associates information from all slots to extract information on spectrum occupancy. Finally, we add learnable position embeddings to these embeddings to form the input embeddings for the subsequent network.

Note that our embedding approach is also applicable to Nyquist-rate sampling, which can be viewed as a special case of multi-coset sampling. In Nyquist-rate sampling, a sequence of data is obtained by a high-speed ADC with a sampling interval of $T$, and within a slot, $L$ samples are taken. This is equivalent to multi-coset sampling with $L$ ADCs with sampling intervals of $LT$. In this case, it is only necessary to adjust the input feature dimension of the FCL for slot embeddings, without requiring any other modifications to the network.
}

\subsubsection{Spectrum Sensing Module}

\name first utilizes Transformer encoders to extract spectrum occupancy features from the embeddings generated by the previous module. The Transformer, which is built upon the self-attention mechanism, has demonstrated remarkable success in NLP tasks and serves as the fundamental component for various large language models. Traditional CNN models extract features locally, making it challenging to capture longer-range features such as frequency domain information from temporal signal inputs (see results in Sec.~\ref{sec:eva:mc}). In contrast, the self-attention mechanism of Transformer enables the model to directly capture long-range dependencies. Furthermore, compared to Recurrent Neural Network (RNN) models that process data sequentially, Transformer is highly parallelizable, allowing for faster training and inference.

For a given sequence of tokens $T \in \mathbb{R}^{N+1, d_{model}}$, the self-attention mechanism uses 3 weight matrices, namely $W_Q$, $W_K$ and $W_V$, with dimensions $d_{model} \times d_{model}$, to transform $T$ into query (Q), key (K) and value (V) matrices. Specifically, $Q=TW_Q$, $K=TW_K$ and $V=TW_V$. Then the output $Z$ of the self-attention layer is
\begin{equation}
    Z = \text{Attention}(Q, K, V) = \text{Softmax}(\frac{QK^T}{\sqrt{d_{model}}})V
    \label{equ:attention}
\end{equation}
where $K^T$ denotes the transpose of $K$ and Softmax is applied element-wise to the resulting matrix. In the self-attention layer, each token in the sequence correlates its corresponding row in the query matrix Q with all rows of the key matrix K, enabling the model to capture correlations with every token.

In practice, the self-attention layer is divided into $h$ parallel headers, where each header employs weight matrices with columns of size $d_{model} / h$. The outputs of these headers are concatenated to form $Z$. Subsequently, $Z$ is linearly projected onto $T'$ through a matrix multiplication with $W$, resulting in $T'$ having the same dimensions as the input $T$. Here, $W \in \mathbb{R}^{N + 1, d_{model}}$.

Following this, a feed-forward network with two FCLs is applied. We use GELU as the activation function in \name. The dimensions of the output from the feed-forward network remain the same as the input. Residual connections and layer normalization are employed after both the self-attention and feed-forward layers. To ensure stable model updates as the model depth increases, DeepNorm~\cite{wang2022deepnet} is utilized at the residual connections. In this approach, the residual is scaled by a factor of $(2N_{layer})^{\frac{1}{2}}$ before being added to the outputs of the self-attention or feed-forward layer. Here, $N_{layer}$ refers to the number of encoders in the module.

\rev{
In \name, we adopt an encoder-only Transformer architecture, excluding the Transformer decoder. This decision is driven by the Transformer encoders' ability to capture bidirectional correlations among slots. 
Unlike the Transformer decoder, which applies masking to limit correlations to a unidirectional context, real-world correlations between slots exhibit bidirectional characteristics.
For example, in the case of a single-carrier signal, the waveform of one symbol can impact the waveforms of preceding and succeeding symbols, leading to inter-symbol interference. In addition, features of an OFDM signal are reflected in the frequency domain, so extracting features in the time domain also requires capturing correlations across longer time duration in both directions. Therefore, in \name, we leverage Transformer encoders without masking.

In the samples embedding module, the input construction involves prepending a [CLS] token to the slot embeddings. This [CLS] token is learned exclusively by the model and is unrelated to the input samples. After passing through several layers of Transformer encoders, the corresponding row vector of this [CLS] token is fed into the spectrum sensor branch, as depicted in Fig.~\ref{fig:network}. The spectrum sensor consists of a single FCL followed by a Sigmoid activation function. Its output is a vector $\hat{s}$ of dimension $n_{band}$. Here the monitored wideband spectrum is divided into $n_{band}$ sub-bands, where each sub-band can accommodate at most one narrowband signal. If $\hat{s}_j > 0.5$, it indicates that the $j$-th sub-band is occupied. This result serves as the output of the spectrum sensing task and is subsequently fed back to the backbone feature extraction network.
}

\subsubsection{Modulation Classification Module}

The monitored wideband spectrum contains multiple signals, and we need to separate them to obtain information of each signal. This module relies on the spectrum sensing results for signal separation. The backbone output features of the spectrum sensing module encompass information about all the signals present within the wideband spectrum. In \name, we replicate these features the same number of times as the occupied sub-bands, and append learnable sub-band identification tokens to the end of each replicated feature. Subsequent networks use these sub-band identification tokens to focus on specific sub-band features.
\rev{
These tokens eliminates the need for additional signal recovery, down-conversion, and filtering processes.
}
In Fig.~\ref{fig:network}, we depict a scenario where the spectrum sensor predicts the occupancy of the $m$-th and $n$-th sub-bands. We duplicate the features and append tokens indicating the $m$-th and $n$-th sub-bands.
We then replace the dirty [CLS] token already used for spectrum sensing with a new and clean modulation [CLS] token, specifically designed for modulation classification. This token is also learnable and is independent of the input samples.

Afterwards, the replicated features are processed independently. The modulation classification module employs Transformer encoders with the same structure as the spectrum sensing module to capture narrowband features. The row vectors corresponding to the modulation [CLS] token are fed into the modulation classifier branch. The modulation classifier consists of a single FCL followed by a Softmax activation function, which will classify the modulation schemes into $n_mod$ classes. The modulation [CLS] token also guides the network to extract symbol-level features, which are subsequently used by the demodulation module.

\subsubsection{Demodulation Module}

Typically, the bandwidth of an individual signal does not exceed several tens of MHz, which is lower than the sampling rate of a common ADC. Therefore, in \name, we require that the bandwidth of sub-band signals does not exceed $1/LT$, which is the sampling rate of a low-speed ADC in multi-coset sampling. This ensures that the number of slots $N$ is not less than the number of symbols carried by a single signal. After being processed by Transformer encoders, while the first token in the features is used for modulation classification, the remaining $N+1$ tokens encode the specific symbol features transmitted by a single signal.

\name consists of $n_{mod}$ demodulators. Based on the results of modulation classification, \name will select the corresponding demodulator. Each demodulator consists of only one FCL with Softmax activation, classifying the remaining $N + 1$ tokens into $M_{mod} + 1$ classes. Here, $M_{mod}$ represents the number of points on the constellation map for modulation scheme $mod$. An extra class is dedicated to the End of Sequence (EOS). The symbols predicted prior to the [EOS] serve as the output of demodulation in \name. This design allows \name to generate symbol sequences of varying lengths for signals with different symbol rates.

\rev{
Note that \name employs the unified backbone to process heterogeneous signals with various modulations, symbol rates, roll-off factors, and other settings. \name use distinct FCLs only when decoding features into symbols. This flexible arrangement enables the model to easily scale with more modulation schemes by incorporating new FCLs.
}

\subsubsection{Loss Functions}

\rev{
For the $k$-th sub-band of the $j$-th input in a training batch, we represent the ground truth spectrum occupation state as $S_{j,k}$. If the $k$-th sub-band is occupied, we set $S_{j,k} = 1$; otherwise $S_{j,k} = 0$. Denote the one-hot coded ground truth modulation as a $n_{mod}$-dimensional vector $m_{j,k}$. Additionally, we use the one-hot vector $y_{j,k,l}$ of dimension $M_{mod} + 1$ to denote the transmitted $l$-th symbol, where the additional dimension represents the [EOS] token. During training, we use the ground truth values $S$ and $m$ to determine the sub-band identification tokens and demodulators. Let $\hat{S}_{j,k}$, $\hat{m}_{j,k}$ and $\hat{y}_{j,k,l}$ denote the corresponding predictions made by \name. Since $\hat{S}_{j,k}$ is activated by the Sigmoid function, and $m_{j,k}$ as well as $y_{j,k,l}$ are activated by Softmax, their values all fall within the range of $(0, 1)$. 

The loss function of \name is designed to address the three signal analysis tasks, which are spectrum sensing, modulation classification and demodulation.

The loss function for spectrum sensing $L_{ss}(S, \hat{S})$ is defined by Eq.~\eqref{equ:loss_ss}. The loss is summed over all $n_{band}$ sub-bands and is then averaged across the batch.
\begin{equation}
\begin{split}
    L_{ss}(S, \hat{S}) = &\frac{1}{N_{batch}}\sum_{j=1}^{N_{batch}}\sum_{k=1}^{n_{band}}w_{j,k}^{\theta}[-S_{j,k}\ln(1-\hat{S}_{j,k}) \\
    &- (1 - S_{j,k})\ln(\hat{S}_{j,k})]
\end{split}
\label{equ:loss_ss}
\end{equation}
Here the exponent $\theta$ is a non-negative parameter and $w_{j,k}$ is a weighting factor employed in focal loss~\cite{lin2017focal} to allow the loss function assign less importance to well-classified samples. $w_{j,k}$ is defined in Eq.~\eqref{equ:loss_ss_w}
\begin{equation}
    w_{j,k} = S_{j,k}(1 - \hat{S}_{j,k}) + (1 - S_{j,k})\hat{S}_{j,k}
    \label{equ:loss_ss_w}
\end{equation}

The loss function for modulation classification, denoted as $L_{mc}(S, M, \hat{M})$, is basically the cross entropy loss, and is defined in Eq.~\eqref{equ:loss_mc}.
\begin{equation}
\begin{split}
    L_{mc}&(S, m, \hat{m}) = \frac{1}{\sum_{j=1}^{N_{batch}}\sum_{k=1}^{n_{band}} S_{j,k}} \\
    & \times \sum_{j=1}^{N_{batch}}\sum_{k=1}^{n_{band}}\sum_{x=1}^{n_{mod}} -m_{j,k,x}\ln(\hat{m}_{j,k,x})S_{j,k}
\end{split}
\label{equ:loss_mc}
\end{equation}
Since modulations of noise on empty sub-bands are not meaningful, we only consider modulations of signals on non-empty sub-bands, as indicated by the term $S_{j,k}$ in Eq.~\eqref{equ:loss_mc}. In practical implementation, we extend the batch by replicating features of non-empty sub-bands, as depicted in Fig.~\ref{fig:network}, while leaving empty sub-bands untouched. Summation over all sub-bands in Eq.~\eqref{equ:loss_mc} is just for mathematical convenience.

We denote $n^{symb}_{j,k}$ as the length of symbols transmitted by the signal in the $k$-th sub-band of the $j$-th input. We label the ground truth $y$ such that $y_{j, k, n^{symb}_{j,k}+1}$ corresponds to the [EOS] token. The loss function for demodulation $L_{demod}(S, y, \hat{y})$ is defined as follows in Eq.~\eqref{equ:loss_demod}.
\begin{equation}
\begin{split}
    L_{demod}&(S, y, \hat{y}) = \frac{-1}{\sum_{j=1}^{N_{batch}}\sum_{k=1}^{n_{band}} S_{j,k}(n^{symb}_{j,k} + 1)} \\
    &\times \sum_{j=1}^{N_{batch}}\sum_{k=1}^{n_{band}}\sum_{l=1}^{n^{symb}_{j,k} + 1}\sum_{x=1}^{M_{mod} + 1} y_{j,k,l,x}\ln(\hat{y}_{j,k,l,x})S_{j,k}
\end{split}
\label{equ:loss_demod}
\end{equation}
Just like the modulation classification loss, only signals in non-empty sub-bands are meaningful. Prediction of [EOS] tokens is also considered in loss function. Finally the total loss function $L(S, m, y, \hat{S}, \hat{m}, \hat{y})$ is defined in Eq.~\eqref{equ:loss} as
\begin{equation}
\begin{split}
    L(S&, m, y, \hat{S}, \hat{m}, \hat{y}) = \\
    &\alpha L_{ss}(S, \hat{S}) + \beta L_{mc}(S, m, \hat{m}) + \gamma L_{demod}(S, y, \hat{y})
\end{split}
\label{equ:loss}
\end{equation}
where $\alpha$, $\beta$, and $\gamma$ are non-negative hyper-parameters to balance the weights for the three tasks.
}

\section{Implementation} \label{sec:implementation}

\begin{figure}[t]
  \centering
  \includegraphics[width=0.9\linewidth]{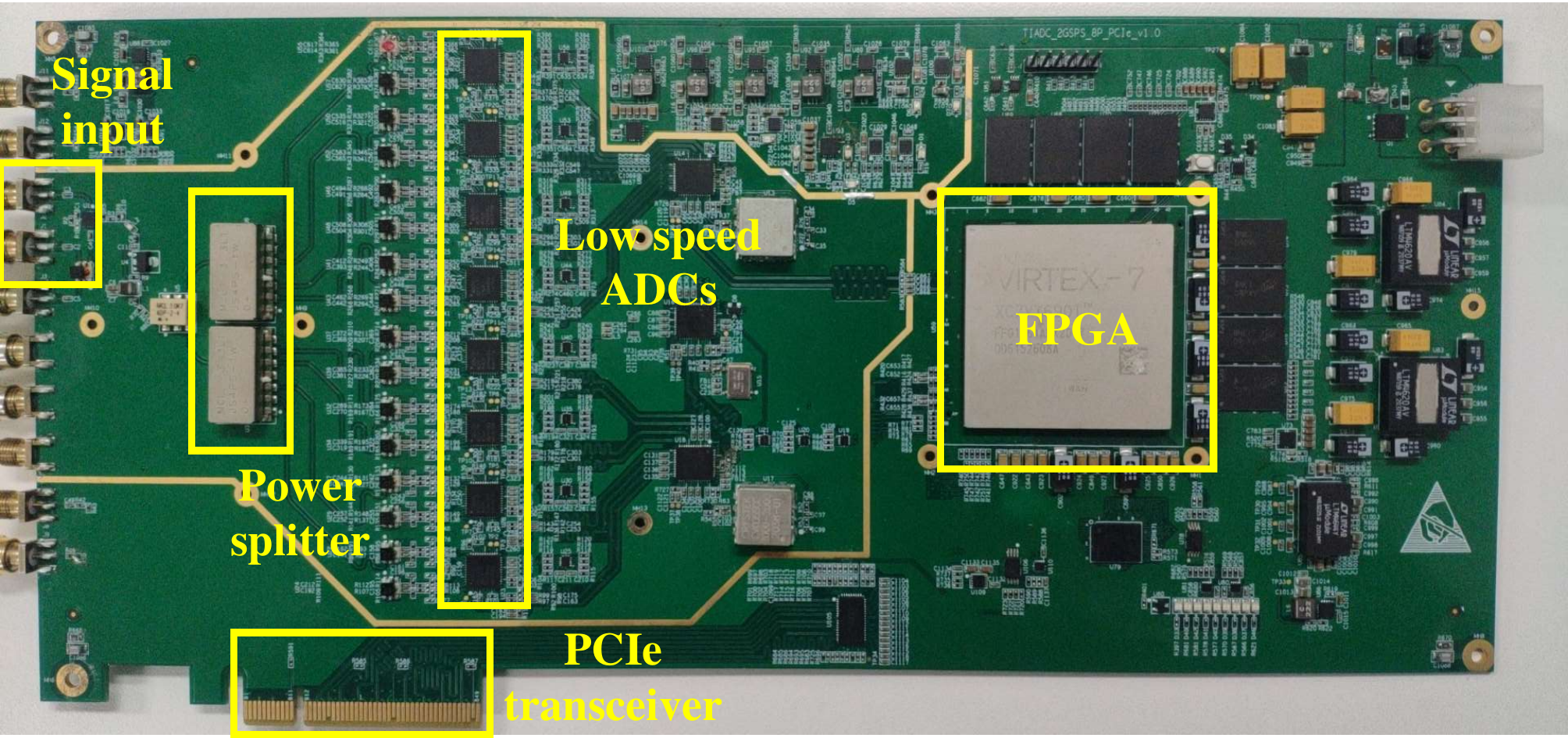}
  \caption{Multi-coset sub-sampling hardware}
  \label{fig:hardware}
  \vspace{-3ex}
\end{figure}

\subsection{Signal sampling}
The transmitting end uses MATLAB to set up signal parameters and Signal Hound VSG60A~\cite{signalHound2023} to get analog signals. The multiband signals we generate follow the signal model described by Eq~\eqref{equ:signal_model}
\begin{equation}
\begin{split}
    r(t) = \sum_j s_j(t) + n(t) = s(t) + n(t)
\end{split}
\label{equ:signal_model}
\end{equation}
In Eq.~\eqref{equ:signal_model}, $s_j(t)$ represents narrowband signals, and $n(t)$ represents the Additive White Gaussian Noise (AWGN). The combined multiband signal $s(t)$ occupies a spectrum ranging from 25\!~MHz to 825\!~MHz, which is equally divided into 16 sub-bands. One or two of the 16 sub-bands are occupied.

A narrowband signal can be either a square root raised cosine shaped single-carrier signal or an OFDM signal. For single-carrier signals, the symbol rate can be either 16\!~MHz or 20\!~MHz, and the roll-off factor can be either 0.05 or 0.25. In case of an OFDM signal, the interval between sub-carriers is 2\!~MHz, and the number of sub-carriers of an OFDM symbol can be 8 or 10.

Signals can be modulated using QPSK, 8PSK, 8QAM or 16QAM. The possible carrier frequencies of a narrowband signal range from 50\!~MHz to 800\!~MHz in intervals of 50MHz. Narrowband signals do not have spectrum collisions. Transmitted bits are randomly generated for each narrowband signal. In order to simulate the near-far effect, the amplitude of the narrowband signals is scaled randomly and independently, such that the signal-to-noise ratio (SNR) for each narrowband signal, denoted as $\text{SNR}_{j} = 10\log_{10}{\frac{<|s_j(t)|^2>}{<|n(t)|^2>}}$, falls within the range of -5\!~dB to 10\!~dB, where $<>$ represents averaging.

At the receiving end, the multi-coset sampling hardware is implemented using 8 parallel ADI AD9250 ADCs~\cite{adiADC2023}. Specifically, the sampling rate for each ADC is 50\!~MSPS. ADC sampling interval to the Nyquist sampling interval ratio $L$ is set to 40. The sampling hardware is shown in Fig.~\ref{fig:hardware}. Signals are segmented into tiny frames with a duration of 0.4\!~$\mu$s. \rev{Ad hoc pilots are added for synchronization.} Samples are embedded as described in Sec.~\ref{sec:system_design:san:embedding} before feeding into the signal analysis network. Training, validation and test data sets are built with the same signal generation settings.


\subsection{Multi-task learning architecture}

We implement the multi-task learning architecture in Python 3.10.11 and PyTorch 1.13. The embedding dimensions denoted as $d_{model}$ in Transformer are set to 256. Each encoder layer utilizes 8 heads for self-attention. The feed-forward layer in the encoder consists of 1024 intermediate features. The default model configuration includes a total of 8 encoders, with 4 dedicated to spectrum sensing and 4 dedicated to demodulation tasks. In order to align the label dimensions to speedup training, we handle the varying number of classes for each demodulator by padding the output vector. Specifically, we pad the output vector from its original dimension $M_{mod} + 1$ to $\max_{mod}M_{mod} + 1$ by appending $-\infty$ values. These padded values do not affect the Softmax activation or symbol prediction. Likewise, in Eq.~\eqref{equ:loss_demod}, the label $y$ is padded with [pad] tokens for alignment and is then filtered to consider only meaningful non-padding tokens for loss calculation.

For the hyper-parameters of the loss function, we set the exponent of the focal loss factor in Eq.~\eqref{equ:loss_ss} as $\theta=2$. The weighting factors for the three signal analysis tasks are set as $\alpha=0.1$, $\beta=0.1$ and $\gamma=1$. Training and evaluations of the signal analysis network are performed on Ubuntu 18.04, with an NVIDIA GeForce RTX 3090 GPU. The network is trained using Adam optimizer with $\beta_1 = 0.9, \beta_2 = 0.999$, a learning rate of 0.001, and a batch size of 1,024. 

\section{Evaluation} \label{sec:evaluation}

\rev{
In the following, we first introduce experiments setup, the oracle-aided guideline, and the compressed sensing benchmark in Sec.~\ref{sec:eva:exp_set}. In Sec.~\ref{sec:eva:ss} we first present the spectrum sensing performance of \name. Sec.~\ref{sec:eva:mc} and Sec.~\ref{sec:eva:phy_recog} present modulation classification and physical layer protocol recognition performance of \name, which shows the design of \name makes the system both accurate and efficient. Sec.~\ref{sec:eva:demod} presents the blind demodulation performance of \name, where our system shows competitive or even better performance compared with oracle-aided traditional methods.
}

\subsection{Experimental Setup} 
\label{sec:eva:exp_set}
We use Simultaneous Orthogonal Matching Pursuit (SOMP)~\cite{tropp2005simultaneous} algorithm as the sub-Nyquist benchmark. SOMP is a spectrum sensing and signal recovery algorithm based on the theory of compressed sensing. Like \name, SOMP takes sub-Nyquist samples as input. It iteratively selects potentially occupied sub-bands and uses a least squares method to reconstruct the signal based on these selections. SOMP terminates when the energy of the remaining sub-bands falls below a certain threshold or the maximum number of iterations is reached. In such cases, the result of the latest least squares estimation is considered as the recovered signal. 
\rev{
The number of low speed ADCs in multi-coset sampling affects the signal analysis performance, so we also evaluate the performance with less sampling ADCs by discarding samples from some channels.
}

For spectrum sensing evaluation, we manually set the optimum stop criterion that yields the highest accuracy. For modulation classification and demodulation tasks, we utilize the SOMP benchmark along with an oracle that supplies actual spectrum occupation states. This allows us to solely employ the least squares method for signal recovery. For demodulation, the oracle further provides prior knowledge such as modulations and symbol rates.

Additionally, we compare the performance of \name with an ideal guideline that samples at a Nyquist rate of 2\!~GSPS. This ideal guideline also benefits from an oracle providing prior knowledge of signal parameters.

We have prepared several model variants by adjusting the depth of the network. The parameters of these model variants are summarized in Tab.~\ref{tab:n_params}. Half of the encoders of a model are used for the spectrum sensing module and half are for the demodulation module. We prepare a training set with 1,000,000 data and a validation set with 10,000 data. The test set for each evaluation contains 10,000 data. All the models are trained once on the training set for 500 epochs. In evaluation, we do not fine-tune the models for specific kinds of signal.

\begin{table}[t]
  \caption{Number of Parameters of Networks}
  \label{tab:n_params}
  \begin{center}
  \begin{tabular}{cc}
    \toprule
    No. of Encoders&No. of Params\\
    \midrule
    4 &3.190\!~M\\
    6 &4.770\!~M\\
    8 &6.350\!~M\\
    10 &7.929\!~M\\
  \bottomrule
\end{tabular}
\end{center}
\vspace{-3ex}
\end{table}

\subsection{Spectrum Sensing} \label{sec:eva:ss}

\begin{figure}[t]
  \centering
  \subfloat[Sampling channels (8 Encs)]{
    \includegraphics[width=0.5\linewidth]{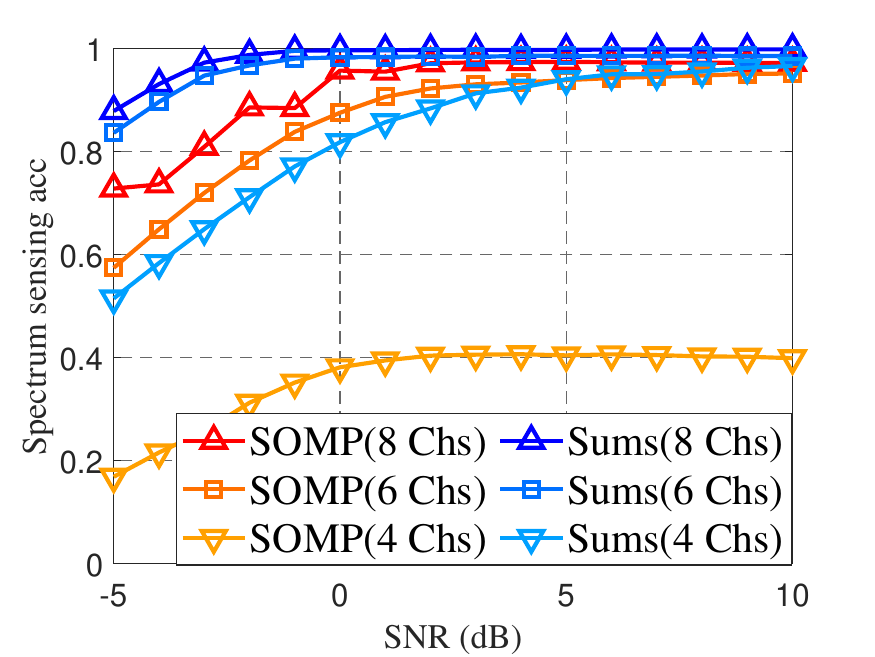}
    \label{fig:ss_chs}
  }
  \subfloat[\name depth (8 Chs)]{
    \includegraphics[width=0.5\linewidth]{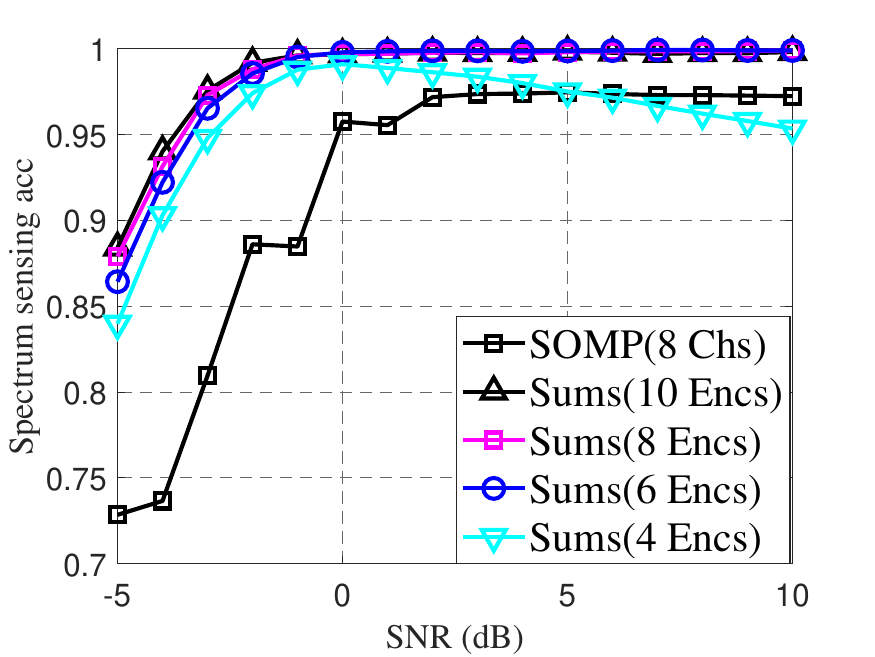}
    \label{fig:ss_encs}
  }
  \caption{Spectrum sensing accuracy under different SNRs.}
  \label{fig:spectrum_sensing}
  \vspace{-3ex}
\end{figure}

In inference, \name relies on the results of the spectrum sensing module to focus on extracting features of the occupied sub-bands. Therefore, spectrum sensing plays an important role in the overall performance of \name. The evaluation results are shown in Fig.~\ref{fig:spectrum_sensing}. A prediction is considered correct only when the occupancy states of all sub-bands are correctly predicted. \rev{SNR in Fig.~\ref{fig:spectrum_sensing} refers to the multiband signal-to-noise ratio $\text{SNR}=10\log_{10}{\frac{<|s(t)|^2>}{<|n(t)|^2>}}$.}

Despite providing the SOMP algorithm with prior information, namely the optimal stop criterion under certain SNR, distinct advantages are still exhibited for different model variants and number of sampling channels. 
\name is always more accurate in spectrum sensing than SOMP optimum under the same number of channels. When only 4 sampling channels are used, \name shows competitive accuracy against SOMP optimal with 6 or 8 channels, whereas the performance of SOMP with 4 channels drops significantly with only accuracy of around 40\% under high SNRs.

\rev{
For the performance with 8 sampling channels, all variants of \name remain accurate under low SNRs. When \name employs only 4 encoders, the accuracy gradually drops as the SNR exceeds 0\!~dB, a trend not observed for \name with more encoders. Most of the errors come from miss-detection, which might be the result of under-fitting.
}

\subsection{Modulation Classification} \label{sec:eva:mc}

\begin{figure}[t]
  \centering
  \subfloat[Confusion matrix of modulation classification by \name]{
    \includegraphics[width=\linewidth]{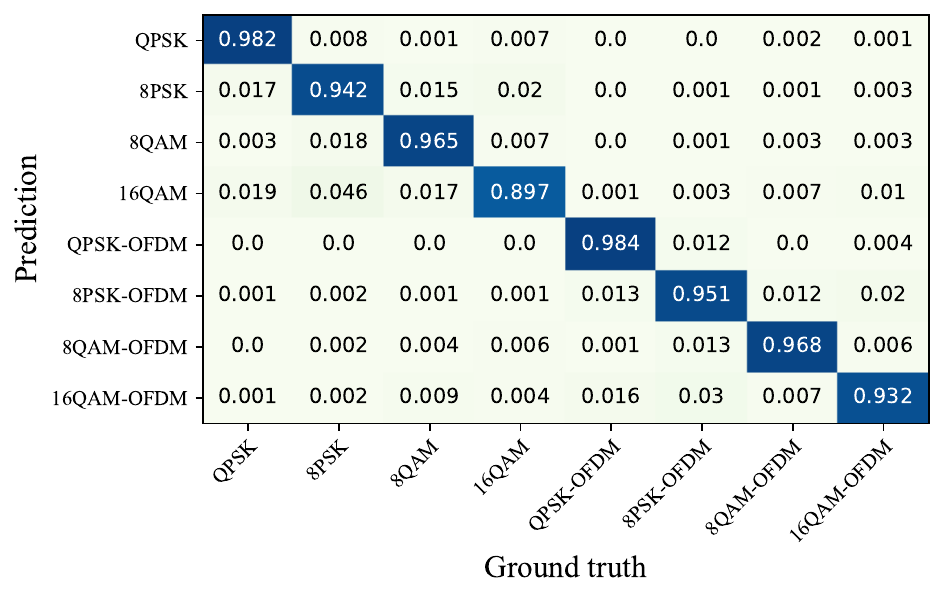}
    \label{fig:confusion_matrix}
  } \\

  \subfloat[Confusion matrix of modulation classification by MCNet with Nyquist sampling]{
    \includegraphics[width=\linewidth]{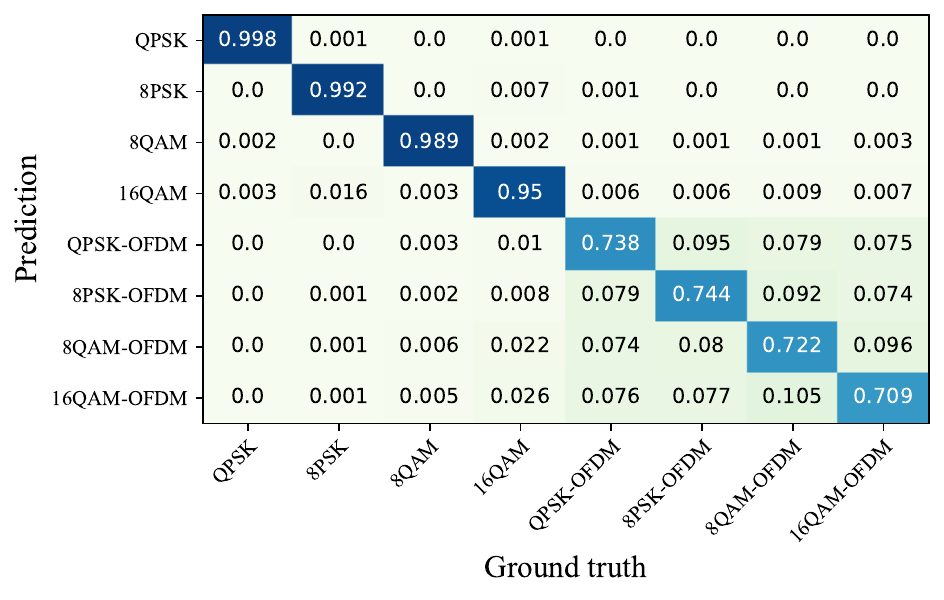}
    \label{fig:confusion_matrix_mcnet_nyq}
  } \\
  
  \subfloat[Sampling channels (8 Encs)]{
    \includegraphics[width=0.5\linewidth]{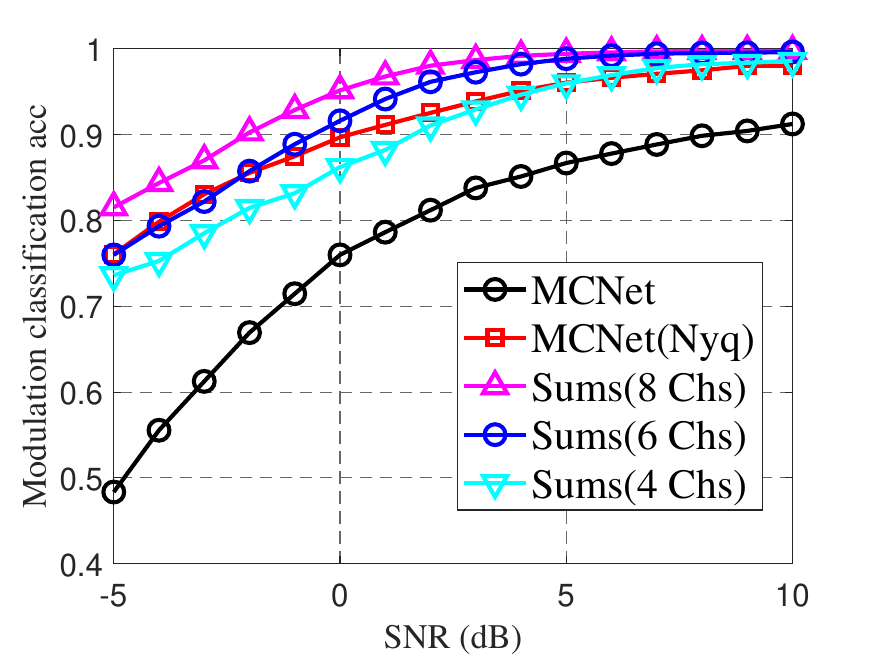}
    \label{fig:amc_chs}
  }
  \subfloat[\name depth (8 Chs)]{
    \includegraphics[width=0.5\linewidth]{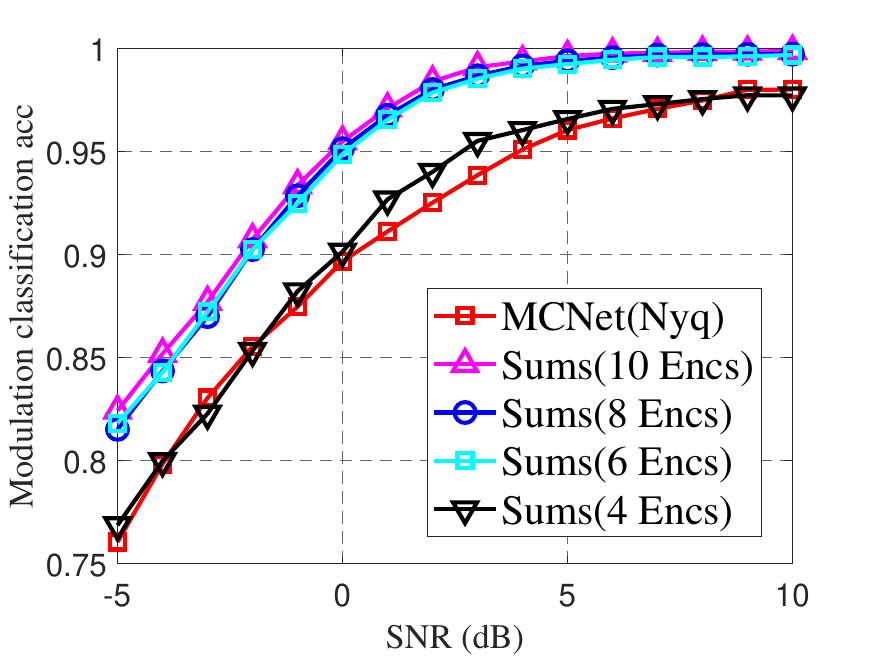}
    \label{fig:amc_encs}
  }
  \caption{\rev{
  Performance of modulation classification. 
  }
  }
  \label{fig:amc}
  \vspace{-3ex}
\end{figure}

In this section, we evaluate the modulation classification performance of \name when the existence of narrowband signals is predicted correctly.  Single-carrier signals and OFDM signals are viewed as differently modulated signals. We compare \name with an open-source benchmark MCNet~\cite{huynh2020mcnet}, which is a ResNet~\cite{he2016deep} and Inception~\cite{szegedy2015going} based CNN model. Input to MCNet is a sequence of samples of a baseband signal. To fit that, we first recover the signal from sub-samples with oracle aided SOMP, then we down-convert and filter narrowband signals into baseband before feeding them to MCNet. In addition, we also filter out narrowband signals from Nyquist samples as the Nyquist benchmark.

Results are shown in Fig.~\ref{fig:amc}. \name is always more accurate than MCNet for sub-samples. Even compared with the Nyquist benchmark, \name shows slight decrease in accuracy only when 4 channels are used. This shows that the signal analysis network can greatly reduce the burden on sampling frontend, while still maintaining competitive modulation classification performance compared with Nyquist-based methods.

Fig.~\ref{fig:confusion_matrix} displays the confusion matrix of \name with 8 encoders and 8 sampling channels, while Fig.~\ref{fig:confusion_matrix_mcnet_nyq} displays the confusion matrix of MCNet with Nyquist input. Both confusion matrices are plotted under an SNR of 0\!~dB. From the confusion matrix of \name, it can be observed that there is a tendency for confusion between 8PSK and 16QAM signals. This is because constellation points of 8PSK belong to the 16QAM constellation as well. With more information from Nyquist sampling, MCNet achieves better accuracy for single-carrier signals. However, MCNet struggles to differentiate OFDM signals effectively, whereas \name exhibits superior performance in this scenario. 

The limitation CNN based models, such as MCNet, lies in its ability to only extract features locally through convolution operations, making it challenging to capture the frequency-domain features from time-domain samples. In contrast, \name leverages the self-attention mechanism of the Transformer, which facilitates the capture of long-range dependencies.

Moreover, despite MCNet has less parameters (180\!~K) compared to \name, in our experimental setup, MCNet requires an average inference time of 0.119\!~ms (in \cite{huynh2020mcnet} 0.125\!~ms) per signal. In contrast, \name with 8 encoders achieves an average inference time of 0.033\!~ms per signal, which includes extra time required for demodulation. This is attributed to the highly parallelizable and non-recursive architecture of the Transformer encoders in \name, and we only need very few samples for analysis without signal recovery.

\rev{
\subsection{Physical Layer Protocol Recognition}
\label{sec:eva:phy_recog}

\begin{figure}[t]
  \centering
  \includegraphics[width=0.7\linewidth]{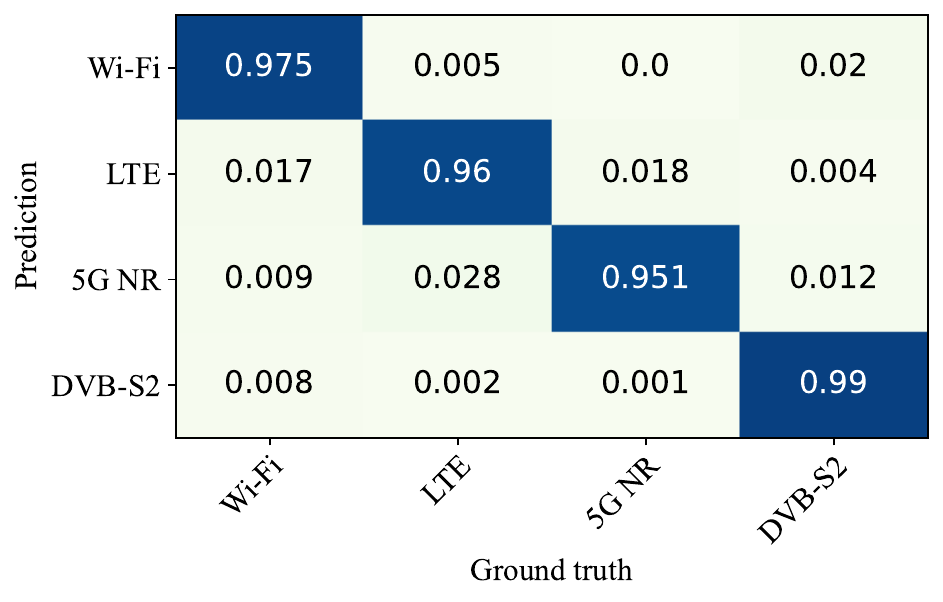}
  \caption{Confusion matrix for physical layer protocol recognition}
  \label{fig:confusion_matrix_heads}
  \vspace{-3ex}
\end{figure}

This section demonstrates the capability of \name in physical layer protocol recognition. We use MATLAB wireless toolboxes to generate multiband signals for various physical layer protocols including Wi-Fi, LTE, 5G NR, and DVB-S2. These signals undergo random Non-line-of-sight fading channels. The modulation classification module of \name is reused for this task. Specifically, we use 3 encoder layers for spectrum sensing, 1 encoder layer for physical layer protocol recognition, and deactivate the demodulation module. We train and validate the model on 10,000 and 2,500 multiband signals, respectively, with SNR ranging from 0\!~dB to 10\!~dB. The confusion matrix obtained from a test dataset comprising 3,000 signals with SNR of -5\!~dB is show in Fig.~\ref{fig:confusion_matrix_heads}, where all 4 types of signals are classified accurately.
}

\subsection{Demodulation Performance}\label{sec:eva:demod}

\begin{figure}[t]
  \centering
  \subfloat[QPSK]{
    \includegraphics[width=0.5\linewidth]{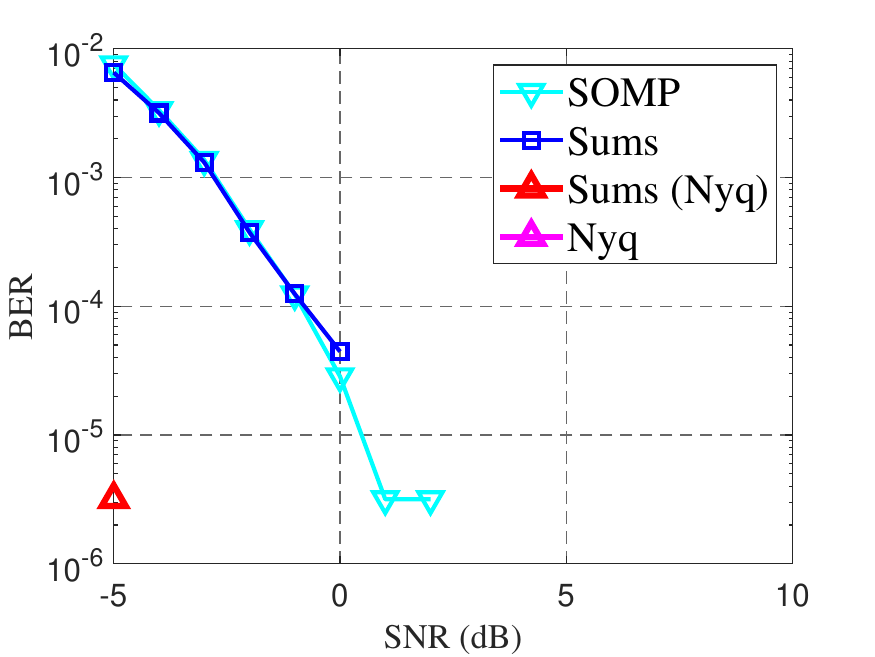}
    \label{fig:qpsk}
  }
  \subfloat[8PSK]{
    \includegraphics[width=0.5\linewidth]{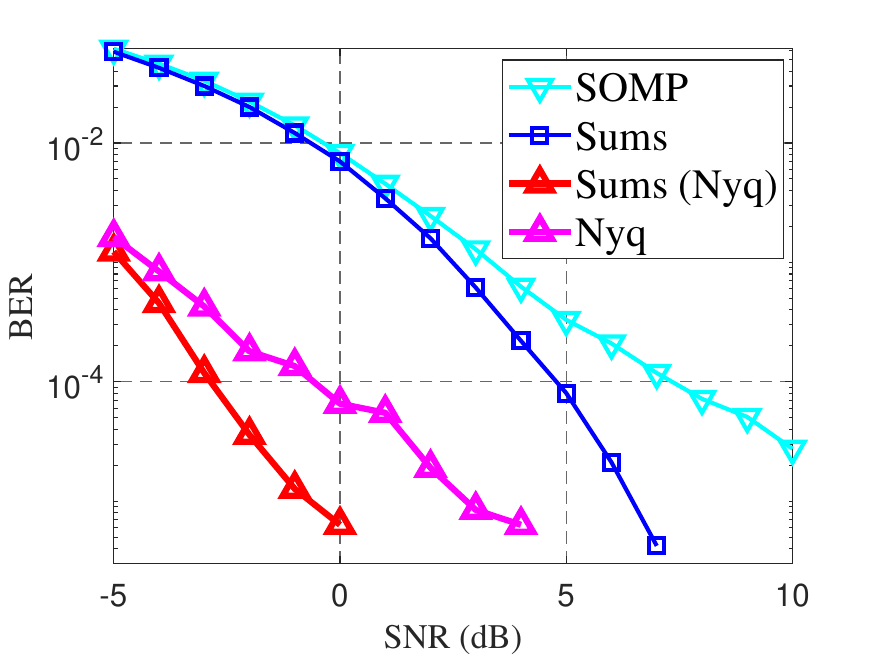}
    \label{fig:8psk}
  } \\
  
  \subfloat[8QAM]{
    \includegraphics[width=0.5\linewidth]{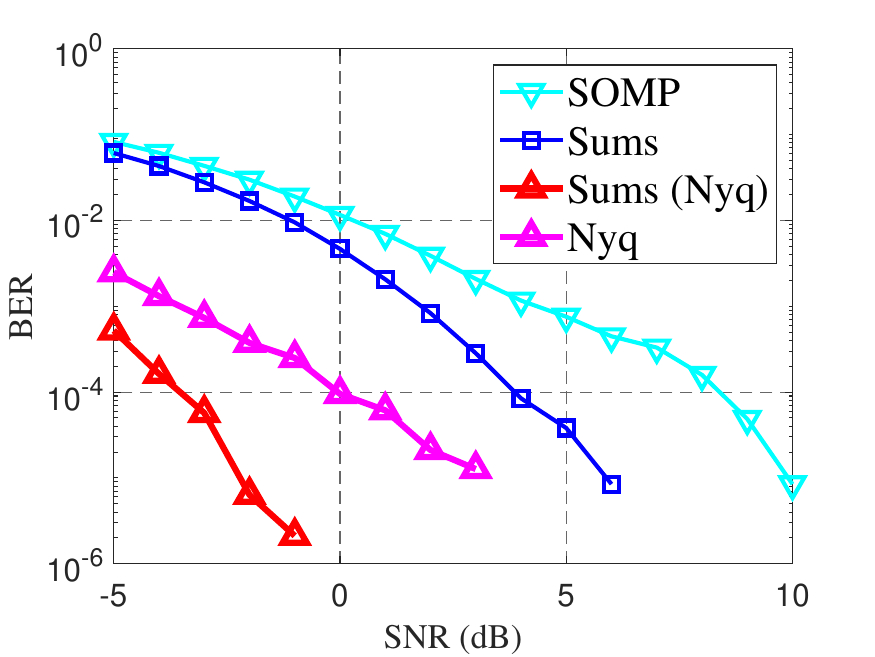}
    \label{fig:8qam}
  }
  \subfloat[16QAM roll-off=0.05]{
    \includegraphics[width=0.5\linewidth]{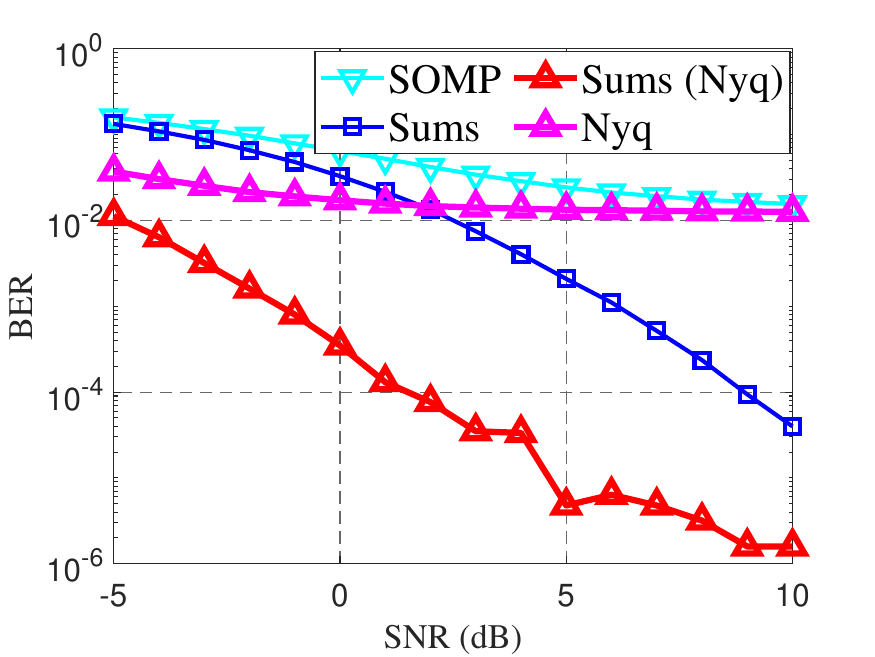}
    \label{fig:16qam_005}
  } \\
  
  \subfloat[16QAM roll-off=0.25]{
    \includegraphics[width=0.5\linewidth]{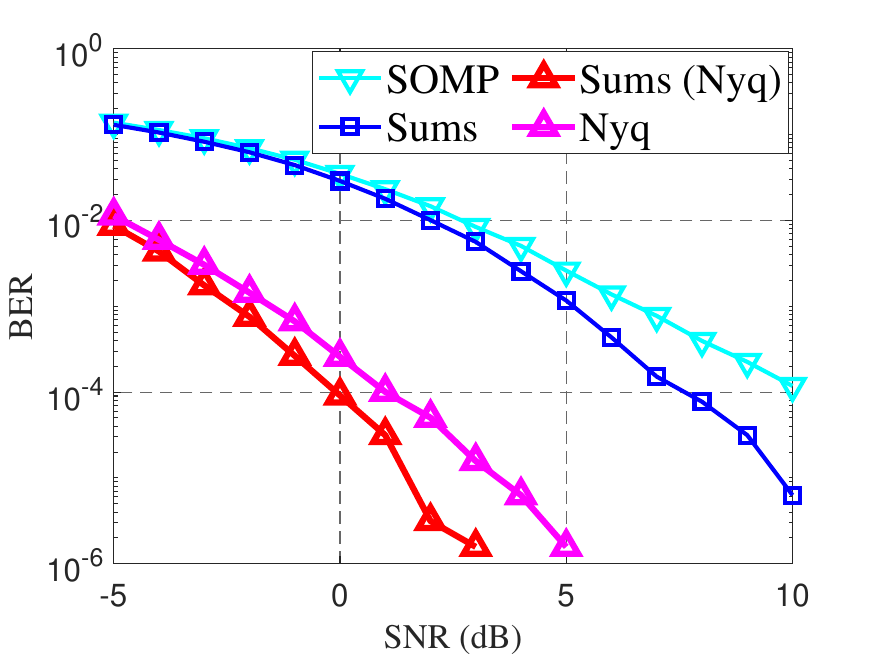}
    \label{fig:16qam_025}
  }
  \subfloat[16QAM-OFDM]{
    \includegraphics[width=0.5\linewidth]{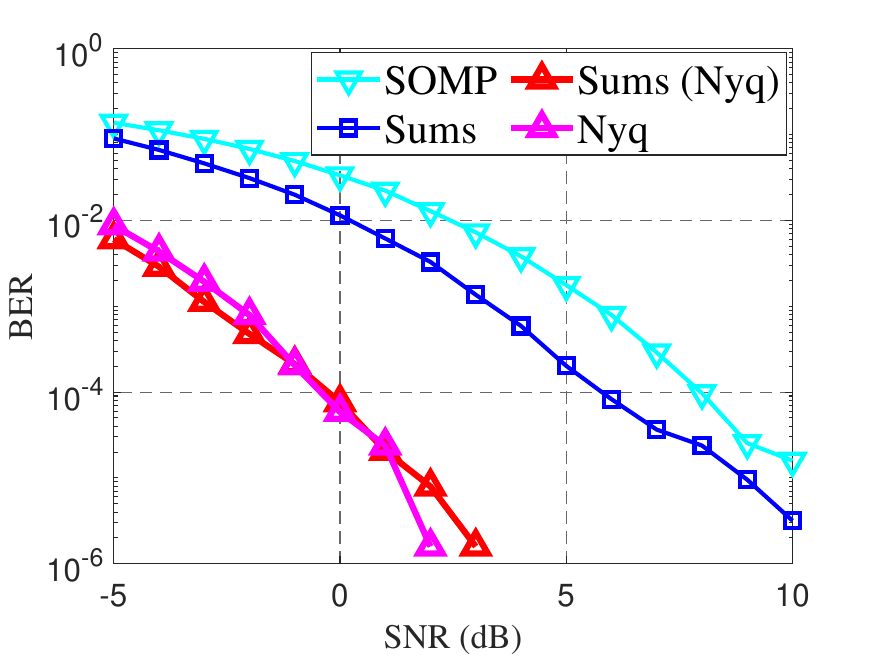}
    \label{fig:16qam_ofdm}
  } \\
  \caption{Demodulation BER}
  \label{fig:demod}
  \vspace{-3ex}
\end{figure}

In this section, we focus on the demodulation performance of \name when the modulations of narrowband signals and symbol lengths are correctly predicted. All other signal parameters are predicted by \name alone. \name in this evaluation use 8 sampling channels and 8 encoders. The performance is compared with the oracle-aided Nyquist guideline and SOMP benchmark that are provided with all prior knowledge except the transmitted bits. The Nyquist guideline performs mixing and match filtering or Fourier Transformation to signals on occupied sub-bands, then demodulate by minimizing Euclidean distance on the constellation map. For the SOMP benchmark, after signal recovery by least square methods, the demodulation process is consistent with the Nyquist guideline. Note that the design of samples embedding module in Sec.~\ref{sec:system_design:san:embedding} enables Nyquist inputs to signal analysis network, so we also evaluate the performance of \name given the Nyquist samples.

The experimental results are shown in Fig.~\ref{fig:demod}.
The SNR-BER relation of multiband signals is different from the SNR-BER relation of baseband signals, where in the multiband scenarios the SNRs are calculated before filter while for baseband signals SNRs are calculated after filter. So here the BERs of multiband signals seem "better" than baseband signals.
Some points are missing from Fig.~\ref{fig:demod} because under these SNRs bits in the test sets are all correctly predicted. For example, there is only one point of \name BER with Nyquist samples in Fig.~\ref{fig:qpsk} and no point of oracle-aided Nyquist BER because most of the signals are rightly demodulated.  In Fig.~\ref{fig:qpsk}, \ref{fig:8psk}, \ref{fig:8qam}, the BER curves of QPSK, 8PSK and 8QAM modulated signals include OFDM signals and square root raised cosine signals with roll-off factors of 0.05 and 0.25. In Fig.~\ref{fig:16qam_005}, \ref{fig:16qam_025}, \ref{fig:16qam_ofdm}, the BER curves of 16QAM modulated signals for three different signal types are displayed separately.

Under the sub-sampling condition, \name consistently outperforms the oracle-aided SOMP across various modulations, since the least square methods for signal recovery aggregate noise across the whole wideband spectrum~\cite{arias2011noise}, while \name is able to focus on specific sub-bands using sub-bands identification tokens. Besides, under the Nyquist sampling condition, \name is still generally superior to the traditional method, even it is aided with an oracle. Oracle aided Nyquist method shows competitive performance against \name only for OFDM signals. This is due to the superiority of \name in correcting inter-symbol interference. 

\rev{
In Fig.~\ref{fig:demod}, it can be observed that \name exhibits more significant advantages over traditional methods in demodulating single-carrier signals. In Fig.~\ref{fig:16qam_005}, when the roll-off factor is only 0.05, the BER curves of both the traditional methods show asymptotic behaviors at high SNRs. Even under sub-sampling conditions, \name achieves performance surpassing that of the traditional Nyquist method. This can be attributed to the fact that signal generators cannot perfectly generate the required shaping pulse. A smaller roll-off factor corresponds to slower pulse attenuation and longer pulse duration, which imposes higher demands on the signal generator, resulting in more severe inter-symbol interference (ISI) caused by imperfect pulse generation. Both the imperfections of the signal generator and the ISI caused by timing misalignment significantly affect the performance of traditional demodulation methods. They struggle to handle ISI effectively since they demodulate individual symbols without considering the influence of neighboring symbols. In contrast, the Transformer encoders of \name capture more inter-symbol dependencies. This compensates for ISI and pulse imperfections, resulting in superior demodulation performance compared to traditional methods.
}

\section{Related Work} \label{sec:related_work}
\textbf{Wideband Spectrum Sensing}. 
The design of \name is inspired by recent developments in spectrum sensing~\cite{arjoune2019comprehensive, kim2008band, chakraborty2017specsense}. Numerous compressed sensing-based methods have been proposed for low-cost wideband spectrum sensing~\cite{hassanieh2014bigband, mishali2009blind, song2019real, mishali2010theory}. Mishali et al. \cite{mishali2009blind} provide theoretical proof for the availability of blind sub-Nyquist spectrum sensing, and propose an alternative sampling scheme called Modulated Wideband Converter (MWC)~\cite{mishali2010theory}. Hassanieh et al. propose BigBand~\cite{hassanieh2014bigband} that leverages sparse Fourier transformation~\cite{hassanieh2012simple} to enable GHz-wide sensing. Deep learning approaches have been considered by treating spectrum sensing as a multi-label classification problem. Nyquist sampling based deep learning approaches~\cite{gao2019deep, uvaydov2021deepsense, selim2017spectrum, davaslioglu2018generative} have been proposed showing better performance over conventional methods. \cite{zhang2022machine} combines compressed sensing and deep learning for spectrum sensing where a multi-coset sampling front end is employed.

\textbf{Blind Communication}.
Several works leverage excellent data feature extraction capability of deep learning for blind communication. Hanna et al.~\cite{hanna2021signal} use several neural networks to estimate signal features separately. Using parameters estimated by neural networks, signals are gradually restored through linear operations and demodulation is made possible. Conventional methods~\cite{swami2000hierarchical, wang2010fast} calculate expert features for automatic modulation classification (AMC) problems. These methods require large number of symbols and are sensitive to noise. O'Shea et al. \cite{o2017introduction} firstly used CNN for AMC, outperforming expert features based methods. Perenda et al~\cite{perenda2021learning, perenda2023contrastive} introduce transfer learning, data augmentation and contrastive learning approaches to address generalization problems of blind estimation neural networks. In addition, multi-task learning approaches are considered for joint signal feature estimation. \cite{wang2021multi, chang2021multitask} adopt multi-task learning framework for AMC. Putra et al.~\cite{putra2022multi} simultaneously perform AMC and packet format detection of WLAN signals, where common features are shared in backbone convolution layers. 

Several works dive deeper into inner information of a signal. Zhang et al.~\cite{zhang2021signal} focus on the classification of Wi-Fi, LTE and 5G NR signal waveforms. IQ samples and their Short-Time Fourier Transformation results are fed into a CNN-LSTM framework. In \cite{wu2019rnn, zheng2022demodnet, zhang2020intelligent, zheng2020deepreceiver}, neural networks show promising performance in demodulation, but these models assume prior knowledge of the signal and a network is only capable to handle one kind of signal. \rev{None of these work provides a system design that can differentiate and process heterogeneous signals simultaneously}

\section{Conclusion} \label{sec:conclusion}

In this paper, we propose a signal analysis system called \name for blind multiband signals analysis in a wideband spectrum. \name benefits from the hardware and algorithm co-design that multi-coset sampling hardware breaks the Nyquist sampling rule to sample more than 1~\!GHz bandwidth with 50~\!~MSPS sampling rate. The compact multi-task learning framework directly processes the heterogeneous sub-Nyquist sampled signal to perform spectrum sensing, modulation classification, physical layer protocol recognition and blind demodulation. Our \name achieves more accurate results than baselines in various signal analysis tasks. We believe \name as a blind monitoring tool can help researchers, operators, and end-users to conveniently realize problems of wireless infrastructures. \rev{Due to the hardware restrictions of the signal generator, we do not evaluate Sums' performance under more complex channel impairments, which we plan to address in the future. We anticipate that such impairments can be easily addressed with the power of deep learning. Further, our future directions include more detailed protocol analysis and cooperative analysis for unknown multiband signals~\cite{li2013maximizing, zeng2010reputation, hao2012hedonic}}.

\ifCLASSOPTIONcompsoc
  \section*{Acknowledgments}
\else
  \section*{Acknowledgment}
\fi


\ifCLASSOPTIONcaptionsoff
  \newpage
\fi



%


\bibliographystyle{IEEEtran}
\bibliography{ref}
\end{document}